\documentclass[aps,prl,nofootinbib,reprint,twocolumn,groupedaddress,altaffilletter,superscriptaddress,showpacs]{revtex4-1}
\usepackage{graphicx}% Include figure filesCM

\usepackage{hep}
\usepackage{relsize}
\usepackage{booktabs}
\usepackage{siunitx}

\newcommand{\ellp}{\ensuremath{\ell^+}\xspace}
\newcommand{\ellm}{\ensuremath{\ell^-}\xspace}

\newcommand{\Dt}{\ensuremath{\Delta t}\xspace}

\newcommand{\chid}{\ensuremath{\chi_d}\xspace}

\newcommand{\ket}[1]{\ensuremath{\vert #1 \rangle}}

\newcommand{\absqop}{\ensuremath{|q/p|}\xspace}

\newcommand{\ACP}{\ensuremath{A_{\CP}}\xspace}

\newcommand{\ifb}{fb\ensuremath{^{-1}}\xspace}
\newcommand{\chisq}{\ensuremath{\chi^2}\xspace}

\newcommand{\on}{{\rm on}}
\newcommand{\off}{{\rm off}}
\newcommand{\cont}{{\rm cont}}
\newcommand{\dir}{{\rm dir}}
\newcommand{\casc}{{\rm casc}}

\mathchardef\Upsilon="7107
\def\Y#1S{\ensuremath{\Upsilon{(#1S)}}\xspace}% no space before {...}!
\def\FourS {\Y4S}
\mathchardef\Lambda="7103
\def\Lbar{\kern 0.2em\overline{\kern -0.2em\Lambda\kern 0.05em}\kern-0.05em{}\xspace}

\newcommand*{\myalign}[2]{\multicolumn{2}{#1}{#2}}

\begin{document}

%%\preprint{\babar-PUB-14/006}
%%\preprint{SLAC-PUB-15983}

{\widetext
\begin{flushleft}
\babar-PUB-14/006\\
SLAC-PUB-15983
\end{flushleft}%
}

%Title of paper
\title{Study of \CP asymmetry in \Bz-\Bzb mixing with inclusive dilepton events}

%% author list as of 06-Jun-2014 (312 authors)
%
\author{J.~P.~Lees}
\author{V.~Poireau}
\author{V.~Tisserand}
\affiliation{Laboratoire d'Annecy-le-Vieux de Physique des Particules (LAPP), Universit\'e de Savoie, CNRS/IN2P3,  F-74941 Annecy-Le-Vieux, France}
\author{E.~Grauges}
\affiliation{Universitat de Barcelona, Facultat de Fisica, Departament ECM, E-08028 Barcelona, Spain }
\author{A.~Palano$^{ab}$ }
\affiliation{INFN Sezione di Bari$^{a}$; Dipartimento di Fisica, Universit\`a di Bari$^{b}$, I-70126 Bari, Italy }
\author{G.~Eigen}
\author{B.~Stugu}
\affiliation{University of Bergen, Institute of Physics, N-5007 Bergen, Norway }
\author{D.~N.~Brown}
\author{L.~T.~Kerth}
\author{Yu.~G.~Kolomensky}
\author{M.~J.~Lee}
\author{G.~Lynch}
\affiliation{Lawrence Berkeley National Laboratory and University of California, Berkeley, California 94720, USA }
% ...
\collaboration{The \babar\ Collaboration}
\noaffiliation

%%\date{\today}

\begin{abstract}
We present a measurement of the asymmetry \ACP between same-sign inclusive
dilepton samples $\ellp \ellp$ and $\ellm \ellm$ ($\ell= e,\, \mu$) from 
semileptonic \B decays in 
$\FourS\to \BB$ events, using the complete data set recorded by
the \babar experiment near the $\FourS$ resonance, 
corresponding to 471 million $\BB$ pairs. 
The asymmetry \ACP  allows comparison between the mixing
probabilities ${\cal P}(\Bzb\to \Bz)$ and ${\cal P}(\Bz\to \Bzb)$,
and therefore probes \CP and \Tsym violation. The result,
$\ACP = (-3.9 \pm 3.5 ({\rm stat.}) \pm 1.9 ({\rm syst.}))\times 10^{-3}$,
is consistent with the Standard Model expectation.

\end{abstract}


\maketitle

A neutral \B meson can transform to its antiparticle through the
weak interaction.
A difference between the probabilities ${\cal P}(\Bzb\ra\Bz)$ and 
${\cal P}(\Bz\ra\Bzb)$ is allowed by the Standard Model (SM), and is
a signature of violations of both \CP and \Tsym symmetries.
This type of \CP violation, called \CP violation in mixing, 
was first observed in the neutral kaon 
system~\cite{Christenson:1964fg,*Dorfan:1967yp,*Bennett:1967zz},
but has not been observed in the
neutral \B system, where the SM predicts an asymmetry of the order of
$10^{-4}$~\cite{Lenz:2010gu,*Lenz:2011ti}. 
The current experimental average of \CP asymmetry in mixing
measured in the \Bz system alone is
$\ACP= (+2.3\pm 2.6)\times 10^{-3}$~\cite{*[{}][{ and online update 
http://www.slac.stanford.edu/xorg/hfag/osc/spring\_2014/.}] Amhis:2012bh},
dominated by the \babar~\cite{Aubert:2006nf,Lees:2013sua}, 
D\O~\cite{Abazov:2012hha}, and Belle~\cite{Nakano:2005jb}
experiments\footnote{
The quoted average excludes the D\O\ inclusive dimuon result~\cite{Abazov:2013uma} and the recently published LHCb result~\cite{Aaij:2014nxa}.}.
A recent measurement in a mixture of \Bz and $B_s^0$ mesons
by the D\O\xspace collaboration deviates from the SM expectation by more than three
standard deviations~\cite{Abazov:2013uma}. Improving the experimental precision 
is crucial for understanding the source of this apparent discrepancy.

The neutral \B meson system can be described by an effective
Hamiltonian ${\mathbf H} = {\mathbf M} - i{\mathbf\Gamma}/2$ for the
two states $\ket{\Bz}$ and $\ket{\Bzb}$. 
Assuming \CPT symmetry, the mass eigenstates can be written as  
$\ket{\B_{L/H}} = p\ket{\Bz} \pm q\ket{\Bzb}$. 
If $|q/p| \neq 1$,
both \CP and \Tsym symmetries are violated.
Details of the formalism can be found in 
Refs.~\cite{Kostelecky:2001ff,Aubert:2004xga}.

The \BBzb pair created in the $\FourS$ decay evolves coherently until
one \B meson decays. In this analysis, we use the charge of the lepton 
(electron or muon) in 
semileptonic \B decays to identify the
flavor of the \B meson at the time of its decay.  
If the second \B meson has oscillated to its antiparticle, it will produce a
lepton that has the same charge as the lepton from the first \B decay.
The \CP asymmetry $\ACP$ between ${\cal P}(\Bzb \to \Bz)$ and 
${\cal P}(\Bz \to \Bzb)$ can be measured by the charge asymmetry of
the same-sign dilepton event rate ${\cal P}_{\ell\ell}^{\pm\pm}$:
\begin{equation}
\ACP =\frac {{\cal P}^{++}_{\ell\ell}-{\cal P}^{--}_{\ell\ell}}{{\cal P}^{++}_{\ell\ell}+ {\cal P}^{--}_{\ell\ell}}
= \frac {1-\absqop^4}{1+\absqop^4}.
\end{equation}
This asymmetry is independent of the \B decay time.

We present herein an updated measurement of \ACP using inclusive 
dilepton events collected by the \babar detector at the
PEP-II asymmetric-energy $e^+e^-$ storage rings at SLAC National 
Accelerator Laboratory. 
The data set consists of $471\times 10^6$ \BB pairs produced at the
$\FourS$ resonance peak (on-peak) and 44~\ifb of data collected 
at a center-of-mass (CM) energy $40$~\MeV below the peak 
(off-peak)~\cite{Lees:2013rw}. 
Monte Carlo (MC) simulated \BB events equivalent to 10 times the data set
based on EvtGen~\cite{Lange:2001uf}
and GEANT4~\cite{Agostinelli:2002hh} with full detector response and
event reconstruction are used to test the analysis procedure.
The main changes with respect to the previous \babar analysis~\cite{Aubert:2006nf}
include doubling the data set, a higher signal selection efficiency, 
improved particle identification
algorithms, and a time-independent approach instead of a time-dependent 
analysis.

The \babar detector is described in detail 
elsewhere~\cite{Aubert:2001tu,*TheBABAR:2013jta}. 
Events are selected if the two highest-momentum particles in the
event are consistent with the electron or muon hypotheses.
All quantities are evaluated in the CM frame unless stated otherwise.
The higher-momentum and lower-momentum lepton candidates are
labeled as 1 and 2, respectively. Four lepton combinations 
are allowed: $\ell_1 \ell_2=\{ee, e\mu, \mu e, \mu\mu\}$,
as are four charge combinations, for a total of 16 subsamples.
We assume $e$-$\mu$ universality, i.e., equal \ACP for all $\ell_1\ell_2$
combinations.
The time-integrated signal yields can be written 
as~\cite{[{See Supplemental Material at [URL will be inserted by publisher] 
for the detailed derivation}]suppl}
\begin{eqnarray}
N_{\ell_1\ell_2}^{\pm\pm} &=& \frac{1}{2}N^0_{\ell_1\ell_2}(1\pm a_{\ell_1}\pm a_{\ell_2}\pm\ACP)\chid^{\ell_1\ell_2}, \label{eq:ssyld}\\
N_{\ell_1\ell_2}^{\pm\mp} &=& \frac{1}{2}N^0_{\ell_1\ell_2}(1\pm a_{\ell_1}\mp a_{\ell_2})(1-\chid^{\ell_1\ell_2}+r_B), \label{eq:osyld}
\end{eqnarray}
in the limit of $\ACP\ll 1$ and  $a_{\ell_j} \ll 1$,
where $a_{\ell_j} = (\epsilon^+_{\ell_j}-\epsilon^-_{\ell_j})/(\epsilon^+_{\ell_j}+\epsilon^-_{\ell_j})$ is the average charge asymmetry of the detection efficiency 
for lepton $j$,
$r_B$ is the \Bp/\Bz event ratio, $\chid^{\ell_1\ell_2}$ is the 
effective mixing probability of neutral \B mesons including efficiency corrections,
and $N^0_{\ell_1\ell_2}$ is the neutral \B signal yield for the $\ell_1\ell_2$
flavor combination.

A small fraction of the background comes from $e^+e^-\to f \bar f (\gamma)$ 
continuum events ($f\in \{u, d, s, c, e, \mu, \tau\}$). This
contribution is subtracted using the off-peak data and the integrated 
luminosity ratio~\cite{Lees:2013rw} between the on-peak and off-peak data sets.
The remaining background comes from \BB events, where at least one lepton
candidate originates from $\B\to X \to \ell Y$ cascade decays, or from
a hadron misidentified as a lepton.

Including the background, we expand 
Eqs.(\ref{eq:ssyld},\ref{eq:osyld}) to parameterize
the total observed numbers of events as
\begin{widetext}
\begin{eqnarray}
M_{\ell_1\ell_2}^{\pm\pm} &=& \frac{1}{2}N^0_{\ell_1\ell_2}(1+R_{\ell_1\ell_2}^{\pm\pm})\Big[ 1\pm a_{\ell_1}\pm a_{\ell_2}\pm 
\frac{1+\delta_{\ell_1\ell_2}R_{\ell_1\ell_2}^{\pm\pm}}{1+R_{\ell_1\ell_2}^{\pm\pm}}\ACP
\Big] \chid^{\ell_1\ell_2}, \label{eq:ssobs}\\
M_{\ell_1\ell_2}^{\pm\mp} &=& \frac{1}{2}N^0_{\ell_1\ell_2}(1+R_{\ell_1\ell_2}^{\pm\mp})(1\pm a_{\ell_1}\mp a_{\ell_2})(1-\chid^{\ell_1\ell_2}+r_B), \label{eq:osobs}
\end{eqnarray}
\end{widetext}
where $R_{\ell_1\ell_2}^{\pm\pm}$ and $R_{\ell_1\ell_2}^{\pm\mp}$ are
background-to-signal ratios under the condition $\ACP=0$, and 
$\delta_{\ell_1\ell_2}$ is the probability of a same-sign background
event being consistent with the flavors of the neutral \B pairs at the time
of their decay after \Bz-\Bzb mixing, i.e., 
$\ell^+\ell^+$ ($\ell^-\ell^-$) for $\Bz\Bz$ ($\Bzb\Bzb$),
minus the probability of the opposite case, i.e.,
$\ell^+\ell^+$ ($\ell^-\ell^-$) for $\Bzb\Bzb$ ($\Bz\Bz$). The detailed
derivation can be found in the supplemental material~\cite{suppl}.
For the opposite-sign events, signal is \CP symmetric. The 
background originating from $\Bz\Bz$ ($\Bzb\Bzb$) preferably contributes
to $\ellp\ellm$ ($\ellm\ellp$) because a primary lepton tends to have
a higher momentum than a cascade lepton. Therefore, the background yield
is also a function of \ACP. However, the coefficient of \ACP is less than 0.01
for the final data sample, so it is ignored in the fits.

Events with $\ge 1$ lepton (single-lepton sample) are used 
to constrain the charge asymmetry of the detector efficiency
$a_{\ell} \equiv (a_{\ell_1}+a_{\ell_2})/2$.
The inclusive single-lepton asymmetry $a_\on$ in on-peak data
can be expressed as~\cite{suppl}
\begin{equation}
a_\on = \alpha + \beta\chid\ACP + \gamma a_\ell,
\label{eq:aon}
\end{equation}
where parameters $\alpha$, $\beta$, and $\gamma$ are functions of 
the following quantities:
the fractions and asymmetries of the continuum background, 
misidentified leptons, and cascade leptons; 
the $\Bz/\Bp$ ratio; and $w^\casc_{\Bz}$ the probability
of the cascade-lepton's charge incorrectly identifying the \B flavor at the time
of the \B decay.

We build a \chisq fit using the 8+8+1 equations represented by 
Eqs.~(\ref{eq:ssobs})--(\ref{eq:aon}) to 
extract \ACP. For the single-lepton sample, we use only electrons 
since the purity is much higher than that of muons.

%%% Event selection %%%
The event selection requires $\ge 4$ charged particle tracks and
the normalized
second-order Fox-Wolfram moment~\cite{Fox:1978vu} $R_2<0.6$. 
The leptons should satisfy $0.6\,\GeV<p_{\ell_2}\le p_{\ell_1}<2.2\,\GeV$.
The polar angle $\theta$ of the electron (muon) candidate in the laboratory
frame is required to satisfy
$-0.788<\cos\theta<0.961$ ($-0.755<\cos\theta<0.956$).
The lepton is rejected if, when combined with
another lepton of opposite charge, the invariant mass is consistent with that 
of a 
$J/\psi$ or a $\psi(2S)$ meson, or when arising from a photon conversion. 
The lepton tracks must pass a set of quality requirements.
For dilepton events, the invariant mass of the lepton pair must
be greater than 150~\MeV. The proper decay time difference \Dt of the two
\B mesons can be determined from the distance along the collision
axis between the points of closest approach of the lepton tracks to 
the beam spot, and the boost factor ($\simeq 0.56$)
of the CM frame. 
We require $|\Dt|<15$~ps and its uncertainty $\sigma_{\Dt}<3$~ps.

Electrons and muons are identified by two separate multivariate algorithms
that predominately use the shower shape and energy deposition in the 
electromagnetic calorimeter 
for electrons, and the track path length and cluster shape in the
instrumented flux return for 
muons.
The electron (muon) identification efficiency is 
approximately 93\% (40\%--80\% depending on momentum). 
The probability of a hadron being identified as an electron (muon) is $< 0.1\%$
($\sim 1\%$).

%%% Multivariate classifier %%%
To further suppress background, we use random forest multivariate 
classifiers~\cite{scikit-learn}. Off-peak data are used to represent
continuum events, and simulated events are used for signal and 
\BB background.
In the dilepton sample, we use six variables: $p_{\ell_1}$, $p_{\ell_2}$, 
thrust and sphericity~\cite{[{See Ch.4 of }]Harrison:1998yr} 
of the rest of the event, the opening angle $\theta_{12}$
of the two tracks in the CM frame, and \Dt.
Separate classifiers are trained on the same-sign and opposite-sign samples.
The $ee$, $e\mu$, $\mu e$, and $\mu\mu$ samples are also trained separately.
The dilepton signal probability distributions of the 
classifiers are shown in 
Fig.~\ref{fig:dilep-rfout}. We select events with a probability 
$>0.7$ to minimize the statistical uncertainty based on fits to the \BB MC
sample. The final on-peak data sample includes $2.5\%$ continuum background
for all dilepton samples,
and 35\% (8\%) \BB background in the same-sign (opposite-sign) sample.

\begin{figure}[htp]
\centering
\includegraphics[width=0.49\textwidth]{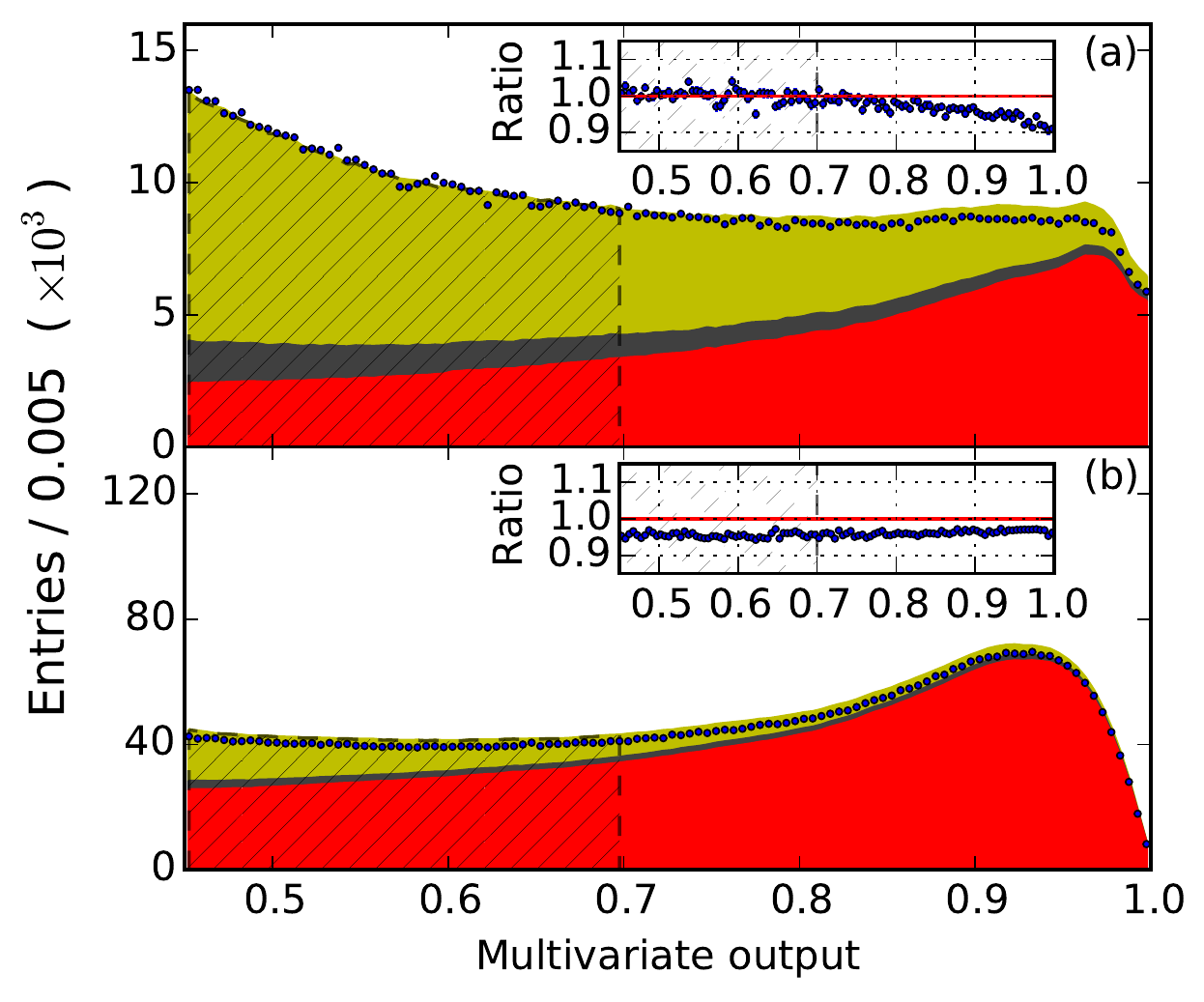}
\caption{(Color online) Signal probability distributions from the 
dilepton multivariate algorithm
for (a) the same-sign sample and (b) the opposite-sign sample; 
all lepton flavors are combined. Points are 
continuum-subtracted data; shaded regions from bottom to top are for 
signal, \BB background with $\ge 1$ misidentified lepton, and \BB background
with both real leptons. Hatched region is rejected. Data/MC ratios are shown in
inset plots. Regions below 0.45 are not shown.}
\label{fig:dilep-rfout}
\end{figure}

% Fake lepton
Approximately 0.1\% (3\%) of selected electrons (muons) in dilepton 
samples are misidentified. 
According to the simulation, nearly 98\% of the misidentified electrons
come from pions and 87\% (12\%) of the misidentified muons come from
pions (kaons).
To correct for the difference in the muon misidentification rates between
data and MC samples, we study the muon identification efficiency
in clean kaon and pion control samples from the
process $\Dst^+\to \Dz\pi^+$ followed by $\Dz\to K^-\pi^+$ (and the
charge-conjugate process). 
The ratios of the efficiencies between data and MC samples
are used to scale the misidentified muon component in the MC sample.
The corrections to $\mu^+$ ($\mu^-$) is $0.792\pm 0.012$ 
($0.797\pm 0.013$).
Since the misidentification rate is very low for electrons, we use a much 
larger pion control sample from $\KS\to\pi^+\pi^-$ decays.
This control sample has a lower
momentum spectrum and does not cover the region of $p>2.5\,\GeV$
in the laboratory frame, which accounts for less than 8\% of the misidentified
leptons. The corrections to misidentified $e^+$ ($e^-$) is $1.00\pm 0.10$ 
($0.56\pm 0.10$). The quoted uncertainties are conservative estimates
that result from mismatched momentum spectra and from a small fraction
of kaons and protons among misidentified electrons.

% Single lepton
For the single-lepton sample, the random forest algorithm uses the 
number of tracks, the event thrust, $R_2$, the difference between the
observed energy in the event and the sum of the $e^+e^-$ beam energies,
the cosines of the angles between the lepton and the axes of the thrust and
the sphericity of the rest of the event, and the zeroth-order and
second-order
polynomial moments $L_0$ and $L_2$, where $L_n = \sum p_i (\cos\theta_i)^n$,
$p_i$ is the momentum of a particles in the rest of the event and 
$\theta_i$ is the angle between that particle and the single-lepton candidate.
We optimize the selection requirement by minimizing the uncertainty of 
the charge asymmetry after the continuum component is subtracted from 
the on-peak data.
A total of $8.5\times 10^7$ single electrons are selected
in the on-peak data, of which approximately 63\% are from direct semileptonic 
\B decays.
Finally, the single electrons are randomly sampled so that the signal
momentum spectrum matches that of the dilepton events.

Raw asymmetries of the single electrons in the on- and off-peak data are 
found to be $a_\on = (4.16\pm 0.14)\times 10^{-3}$ and 
$a_\off = (11.1\pm 1.4)\times 10^{-3}$. The larger asymmetry in the 
off-peak data is primarily due to the radiative Bhabha background 
and the larger detector 
acceptance in the backward (positron-beam) direction. 
The continuum fraction $f_\cont = (10.32\pm0.02)\%$ 
is obtained from the ratio of the selected single 
electrons and the integrated luminosities in off- and on-peak 
data~\cite{Lees:2013rw}.
The neutral \B fraction in the \BB component $f_{\Bz}= (48.5\pm 0.6)\%$ 
is the $\FourS\to\BBzb$ branching fraction~\cite{Beringer:1900zz}
corrected for the selection efficiency.
The cascade event fractions $f_{\Bz}^\casc= 19.8\%$ and $f_{\Bpm}^\casc= 15.3\%$
are obtained from simulation, with negligible statistical uncertainties.
The fraction of the misidentified electron is 0.19\%, and the
asymmetry is approximately 35\%.
The difference between direct and cascade electron asymmetries is 
 $(-1.16\pm 0.25)\times 10^{-3}$ in MC. 
The probability $w^\casc_{\Bz}$ in MC is found to be $(73.8\pm 0.1)\%$.
Using these numerical values, we determine the coefficients in 
Eq.~(\ref{eq:aon}):
$a_\on-\alpha = (2.60\pm 0.20)\times 10^{-3}$, $\beta\chid = 0.057\pm0.001$,
and $\gamma = 0.8951\pm0.0002$.

%%%%%% chisq fit  %%%%%

The fitting procedure is tested on the \BB MC sample; the result 
$\ACP^{\rm MC}= (-1.00\pm 1.04)\times 10^{-3}$ is consistent with the 
\CP-symmetric simulation model. 
We artificially create a non-zero \ACP by reweighing mixed events
in the MC sample, and confirm that the fitting procedure tracks the change
in the \ACP without bias.
The continuum-subtracted event yields are shown in Table~\ref{tab:16signalyields}
and are used in Eqs.~(\ref{eq:ssobs}--\ref{eq:osobs}) for the fit.
The result of the fit to data, after correcting for the small bias
($-1.0\times 10^{-3}$) in the simulation, is $\ACP = (-3.9 \pm 3.5)\times 10^{-3}$,
$a_{e_1}=(3.4\pm0.6)\times 10^{-3}$, $a_{e_2}=(3.0\pm0.6)\times 10^{-3}$, 
$a_{\mu_1}=(-5.6\pm1.1)\times 10^{-3}$, and 
$a_{\mu_2}=(-6.5\pm1.1)\times 10^{-3}$.
The remaining free parameters are $N^0_{\ell_1\ell_2}$ and $\chid^{\ell_1\ell_2}$. 
The \chisq value is 6.2 for 4 degrees of freedom.
The correlations between \ACP and $a_{e_1}$, $a_{e_2}$, $a_{\mu_1}$, 
and $a_{\mu_2}$ are $-0.41$, $-0.47$, $-0.54$, and $-0.51$, respectively. 
Correlations among other parameters are negligible. 
Figure~\ref{fig:subsamples} shows the fit results for the six data-taking
periods and the four flavor subsamples.

\begin{table}
\caption{Continuum-subtracted number of events.}
\centering
\begin{tabular*}{0.48\textwidth}{@{\extracolsep{\fill}}
c
S[table-format=5.0]@{\,\( \pm \)}
S[table-format=3.0]
S[table-format=6.0]@{\,\( \pm \)}
S[table-format=3.0]
S[table-format=6.0]@{\,\( \pm \)}
S[table-format=3.0]
S[table-format=5.0]@{\,\( \pm \)}
S[table-format=3.0]
}
\hline\hline
   & \myalign{c}{$\ellp\ellp$} & \myalign{c}{$\ellp\ellm$} & \myalign{c}{$\ellm\ellp$} & \myalign{c}{$\ellm\ellm$} \\
\hline
$ee$ & 82303 & 320 & 426296 & 783 & 425309 & 782 & 81586 & 323 \\
$e\mu$ & 55277 & 263 & 384552 & 684 & 378261 & 660 & 55878 & 264 \\
$\mu e$ & 67399 & 290 & 467591 & 737 & 475363 & 744 & 67152 & 290 \\
$\mu\mu$ & 47384 & 243 & 277936 & 619 & 278691 & 618 & 48145 & 247 \\
\hline
\end{tabular*}
\label{tab:16signalyields}
\end{table}%

\begin{figure}[htp]
\centering
\includegraphics[width=0.48\textwidth]{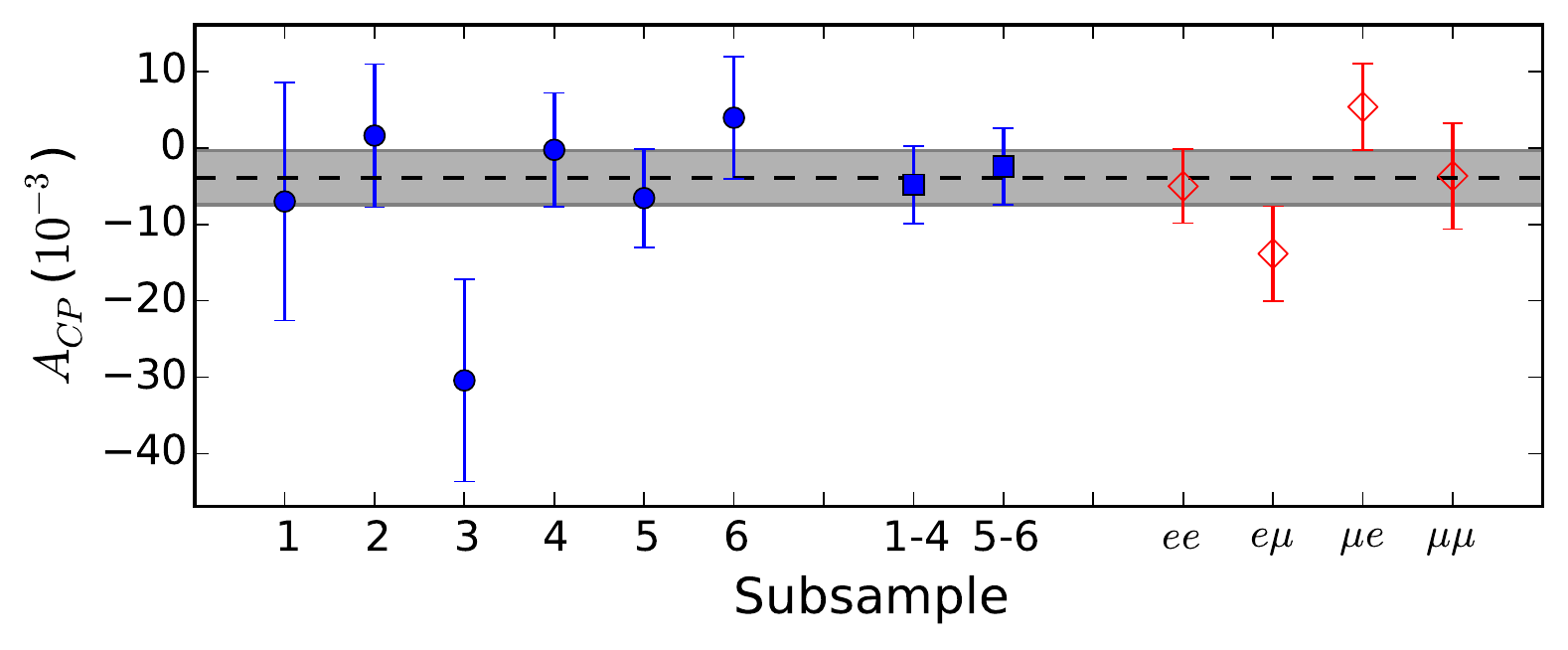}
\caption{(Color online) \ACP of the six data-taking periods (dots), 
the first four and the last two periods (squares), and the four flavor 
subsamples (rhombuses).
The horizontal band is the $\pm1\sigma$ region of the final fit result.
All error bars are statistical only.}
\label{fig:subsamples}
\end{figure}

%% Systematics %%%

The systematic uncertainties are summarized in Table~\ref{tab:systematics}.
The branching fractions in the \B decay chain partially
determine the background-to-signal ratio. 
We correct the MC samples so that important branching fractions
are consistent with the world average~\cite{Beringer:1900zz}. 
These branching fractions correspond to inclusive \B semileptonic decays, 
$\B\to\tau\nu_\tau X$, charm production 
(\Dz, \Dzb, \Dpm, $\D_s^{\pm}$, $\Lambda_c^+$, and $\Lbar_c^-$) 
from \B decays, and inclusive charm semileptonic
decays. The corrections vary for most decays between 0.57 and 1.32, depending
on the channel. We estimate the systematic uncertainty by varying the 
corrections over their uncertainties, which are dominated by the errors
of the world averages.

The systematic uncertainties due to misidentified leptons 
are estimated by varying the uncertainties of the
corrections to $e^+$, $e^-$, $\mu^+$, and $\mu^-$ individually,
and separately for the dilepton and single-electron samples.

In the single-electron MC sample, the charge asymmetry of the electron
in \BBzb is slightly different from that in \BBpm by 
$(0.46\pm0.18)\times 10^{-3}$.
Since we cannot separate \BBpm electrons from \BBzb electrons in data, the 
single-electron asymmetry measurement is the average of the two asymmetries,
which is half the difference away from the \BBzb electron charge
asymmetry.
The systematic uncertainty is determined by the change in \ACP after
shifting the asymmetry in the signal component of the single-electron
sample by half the charge asymmetry difference .

The difference in charge asymmetry between the direct and the cascade electrons
is found to be $a_e^\casc-a_e^\dir = (-1.16\pm 0.25)\times 10^{-3}$ in
the single-electron MC sample.
The difference between the lower-momentum and the higher-momentum electron asymmetries
is negative. This trend is consistent with the result of the fit to
the dilepton data: 
$a_{e_2}-a_{e_1}= (-0.4\pm 0.7)\times 10^{-3}$. For muons, the corresponding
values are $a_\mu^\casc-a_\mu^\dir = (-0.47\pm 0.28)\times 10^{-3}$ and 
$a_{\mu_2}-a_{\mu_1}= (-0.9\pm 1.2)\times 10^{-3}$.
In each case, we set the cascade lepton charge asymmetry to that of
the direct lepton, and use the change in \ACP as a systematic uncertainty.

The background-to-signal ratios $R_{\ell_1\ell_2}^{\pm\pm}$ and 
$R_{\ell_1\ell_2}^{\pm\mp}$ (under the condition \ACP=0) in the
dilepton sample are determined 
from the MC sample. The correction for the misidentified lepton background
has been dealt with above. The real lepton portion of the ratio is in 
principle the same between $\ellp\ellp$ and $\ellm\ellm$ samples because
the particle identification efficiencies cancel between the background
and the signal. In the MC sample, they are consistent within 1\,$\sigma$.
Varying $R_{\ell_1\ell_2}^{++}$ and $R_{\ell_1\ell_2}^{--}$ or
$R_{\ell_1\ell_2}^{+-}$ and $R_{\ell_1\ell_2}^{-+}$
simultaneously in the same direction results in negligible changes in \ACP.
If they are varied independently, the quadratic sum of the changes in \ACP 
is larger. We use the latter as a systematic uncertainty.

The random forest output distribution in the data could be different from
that in the MC sample. The selection efficiency in the MC \BB dilepton
events is approximately 2\% larger than that in the data.
We move the dilepton random forest selection for
the MC sample, while keeping data the same, so that the selected MC events
are reduced by up to 6\%. We take the average change in \ACP as a 
systematic uncertainty.

Several other sources of systematic uncertainties are studied and 
found to be negligible. These include the overall dilepton signal fraction 
estimate, the kinematic difference between on-peak and off-peak
data due to different CM energies, the continuum component fraction, 
the probability $w^\casc_{\Bz}$,
the neutral-to-charged \B ratio, the same-sign background dilution factors
$\delta_{\ell_1\ell_2}$, and the overall cascade event fraction.

\begin{table}
\caption{Summary of systematic uncertainties on \ACP.}
\centering
\begin{tabular*}{0.47\textwidth}{l@{\extracolsep{\fill}}c}
\hline\hline
Source & $(10^{-3})$ \\
\hline
Generic MC bias correction                 & 1.04 \\
MC branching fractions                     & 0.43 \\
Misidentified lepton corrections in dilepton events & 0.77 \\
Misidentified $e$ correction in single electron events    & 0.65 \\
Neutral/charged \B difference              & 0.74 \\
Direct-/cascade $e$ asymmetry difference   & 0.44 \\
Direct-/cascade $\mu$ asymmetry difference & 0.34 \\
Background-to-signal ratios                & 0.68 \\
Random forest cut efficiency               & 0.08 \\
\hline
Total                                      & 1.90 \\
\hline\hline
\end{tabular*}
\label{tab:systematics}
\end{table}%

%% Conclusion %%
In conclusion, we measure the \CP asymmetry 
$\ACP = (-3.9 \pm 3.5\pm 1.9)\times 10^{-3}$
in \Bz-\Bzb mixing using inclusive dilepton decays.
This result is consistent with the SM prediction and the world
average~\cite{Amhis:2012bh}.
This measurement represents a significant improvement with respect to our
previous result~\cite{Aubert:2006nf} (superseded by this result), and is
among the most precise measurements~\cite{Beringer:1900zz,Aaij:2014nxa}.
A Comparison of experimental results
and averages is shown in Fig.~\ref{fig:adas2d}.

\begin{figure}[htbp]
\centering
\includegraphics[width=0.45\textwidth]{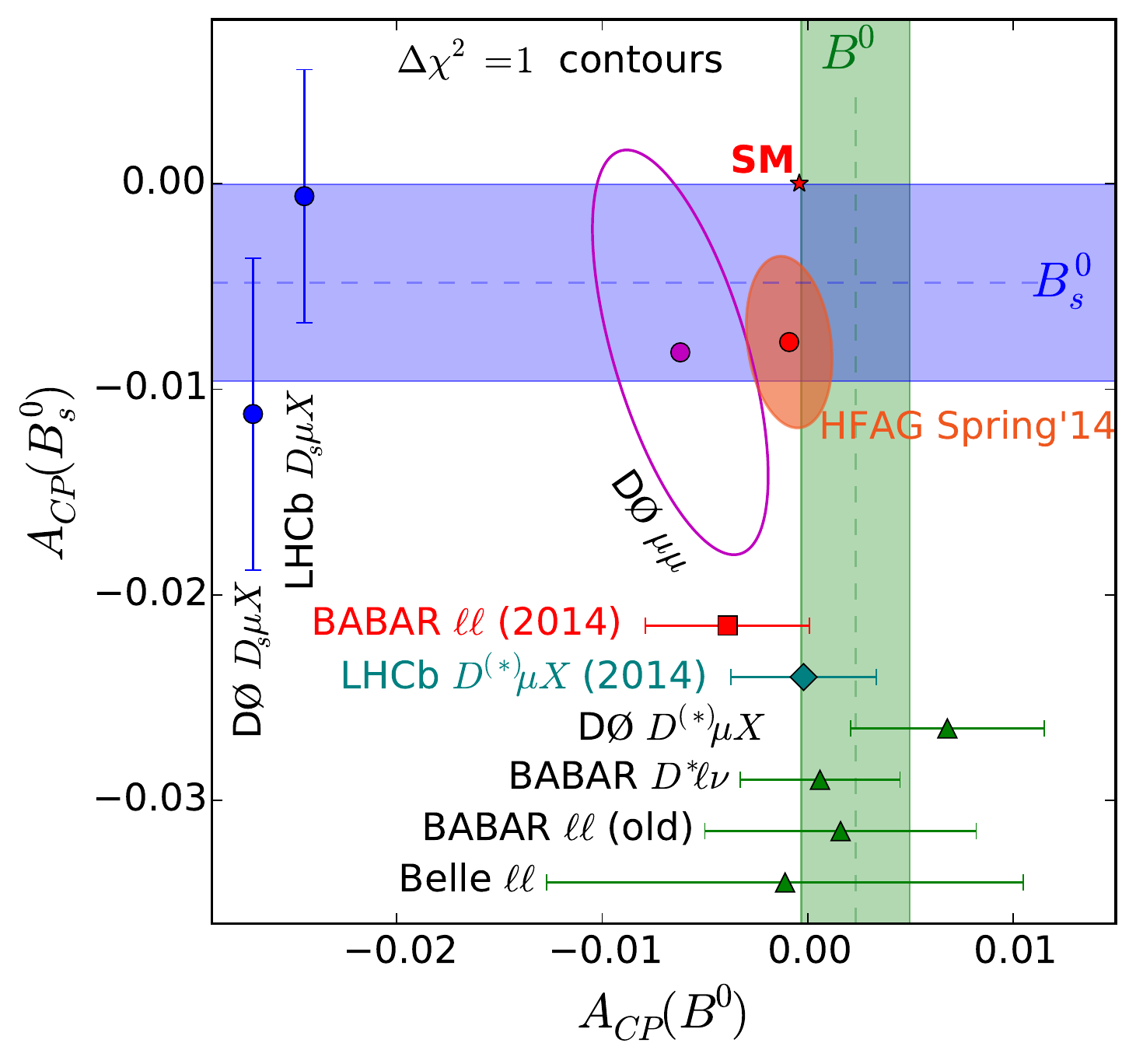}
\caption{(Color online) 
Measurements of \CP asymmetry in neutral \B mixing, including this 
measurement (red square), recent LHCb result~\cite{Aaij:2014nxa} (teal rhombus),
Refs.~\cite{Aubert:2006nf,Lees:2013sua,Abazov:2012hha,Nakano:2005jb} 
(\Bz; green triangles),
Refs.~\cite{Abazov:2012zz,Aaij:2013gta} (\Bzs; blue dots), 
and Ref.~\cite{Abazov:2013uma} (\Bz, \Bzs mixture; magenta contour).
The vertical band is the average of 
Refs.~\cite{Aubert:2006nf,Lees:2013sua,Abazov:2012hha,Nakano:2005jb} and several
other older measurements (not shown). The horizontal band is the average of Refs.~\cite{Abazov:2012zz,Aaij:2013gta}. The world average ``HFAG Spring '14''~\cite{Amhis:2012bh} is also shown (orange contour).}
\label{fig:adas2d}
\end{figure}

%% acknowledgement
We are grateful for the excellent luminosity and machine conditions
provided by our PEP-II colleagues, 
and for the substantial dedicated effort from
the computing organizations that support \babar.
The collaborating institutions wish to thank 
SLAC for its support and kind hospitality. 
This work is supported by
DOE
and NSF (USA),
NSERC (Canada),
CEA and
CNRS-IN2P3
(France),
BMBF and DFG
(Germany),
INFN (Italy),
FOM (The Netherlands),
NFR (Norway),
MES (Russia),
MINECO (Spain),
STFC (United Kingdom),
BSF (USA-Israel). 
Individuals have received support from the
Marie Curie EIF (European Union)
and the A.~P.~Sloan Foundation (USA).

% Create the reference section using BibTeX:
\bibliography{paper}

\end{document}